\documentclass[aps,11pt, superscriptaddress, showpacs,showkeys,floatfix]{revtex4}

\usepackage{graphicx}

\newcommand{\mycaption}[1]{\caption{\footnotesize {#1}}}

\begin{document}

\title{Quantization of Friedmann-Robertson-Walker spacetimes in the
presence of a negative cosmological constant and radiation}

\author{G. A. Monerat\footnote{E-mail: monerat@uerj.br}, 
E. V. Corr\^{e}a Silva\footnote{E-mail: evasquez@uerj.br}, 
G. Oliveira-Neto\footnote{E-mail: gilneto@fat.uerj.br}}

\address{Departamento de Matem\'{a}tica e Computa\c{c}\~{a}o, 
Faculdade de Tecnologia, \\ 
Universidade do Estado do Rio de Janeiro, Estrada Resende-Riachuelo, s/n$^o$, Morada da Colina \\
CEP 27523-000, Resende-RJ, Brazil.}

\author{L. G. Ferreira Filho\footnote{E-mail: gonzaga@fat.uerj.br}}

\address{Departamento de Mec\^{a}nica e Energia, 
Faculdade de Tecnologia,\\ 
Universidade do Estado do Rio de Janeiro, Estrada Resende-Riachuelo, s/n$^o$, Morada da Colina \\
CEP 27523-000 , Resende-RJ, Brazil.}

\author{N. A. Lemos\footnote{E-mail: nivaldo@if.uff.br}}

\address{Instituto de F\'{\i}sica, Universidade Federal Fluminense, \\
R. Gal. Milton Tavares de Souza s/n$^o$, Boa Viagem\\
CEP 24210-340, Niter\'oi-RJ, Brazil}

\date{\today}

\begin{abstract}
In the present work, we quantize
three Friedmann-Robertson-Walker models in the presence 
of a negative cosmological constant and radiation. The models differ from 
each other by the constant curvature of the spatial sections, which may be 
positive, negative or zero. They give rise to Wheeler-DeWitt equations 
for the scale factor which have the form of the Schr\"{o}dinger equation 
for the quartic anharmonic oscillator. We find their eigenvalues and 
eigenfunctions by using a method first developed by 
Chhajlany and Malnev. After that, we use the eigenfunctions in 
order to construct wave packets for each case and evaluate the 
time-dependent expected value of the scale factors. We find for all of them
that the expected values of the scale factors oscillate between 
maximum and minimum values. Since the expectation values of the scale 
factors never vanish, we conclude that these models do not have singularities.
\end{abstract}



\maketitle

\section{Introduction} 

One of the motivations for the quantization of cosmological models was
that of avoiding the initial {\it Big Bang} singularity. 
Since the pioneering work in quantum cosmology due to DeWitt \cite{dewitt},
workers in this field have been attempting to prove
that quantum cosmological models entail only
regular space-times. An important contribution to this issue was
given by Hartle and Hawking \cite{hawking}, who proposed the {\it
no-boundary} boundary condition, which selects only 
regular space-times to contribute to the wave-function of the
Universe, derived in the path integral formalism. Therefore, by
construction, the {\it no-boundary} wave-functions are everywhere
regular and predict a non-singular initial state for the Universe.
Using that boundary condition, in certain particular
cases the {\it no-boundary} wave-function can be explicitly 
computed \cite{hawking, gil, fujiwara}. Another way by 
which one may
compute the wave-function of the Universe is by directly solving the 
Wheeler-DeWitt equation \cite{dewitt}. The wave-function of the Universe 
for some important models have been computed using this 
approach \cite{moss, lemos}. We should mention another boundary 
condition, the {\it tunelling} boundary condition, proposed by Vilenkin 
\cite{vilenkin}, to be imposed on solutions to the Wheeler-DeWitt 
equations. The {\it tunelling} wave-function was also shown to give
rise to models free from the initial {\it Big Bang} singularity 
\cite{vilenkin}. More recently, the absence of singularities in quantum 
cosmological models has been investigated by using the
DeBroglie-Bohm interpretation of quantum mechanics \cite{bohm}. In this
interpretation, one may compute the dynamical trajectories for the 
quantum variables of the system. In particular, since for most of the
quantum cosmological models one uses minisuperspaces \cite{misner}, one
computes, through the DeBroglie-Bohm interpretation, the dynamical 
trajectories for the scale factor. Then, for the great majority of cases
studied so far, the scale factor never vanishes, which implies that
the model has no initial singularities \cite{germano, germano1}.
Although we shall restrict our attention to a model in quantum general 
relativity, it is important to mention that several works in loop
quantum cosmology have also shown that the wave-function of the Universe
is free from initial singularities \cite{bojowald, bojowald1}.

Several important theoretical results and predictions in quantum cosmology
have been obtained with a negative cosmological constant. Considering
a subset of all four-dimensional spacetimes with constant negative curvature 
and compact space-like hypersurfaces, Carlip and coworkers showed how to 
compute the sum over topologies leading to the {\it no-boundary} wave-function 
\cite{carlip, carlip1}. These spacetimes are curved only due to the
presence of a negative cosmological constant. In Ref. \cite{carlip} it was shown
how to obtain a vanishing cosmological constant as a prediction from the
{\it no-boundary} wave-function and in Ref. \cite{carlip1} it was shown how to
obtain predictions about the topology of the Universe from the {\it no-boundary} 
wave-function. We may also mention the result in Ref. \cite{gil}, where the
WKB {\it no-boundary} wave-function of a homogeneous and isotropic Universe with
a negative cosmological constant was computed. Due to the regularity condition
imposed upon the space-times contributing to the {\it no-boundary}
wave-function, it was shown that only a well defined, discrete spectrum 
for the cosmological constant is possible. It was also found that among 
the space-times contributing to wave function, there were two complex
conjugate ones that showed a new type of signature change.
 
It is important to mention that although recent observations point toward a
positive cosmological constant, it is still possible that at the very early
Universe the cosmological constant be negative. Besides that, we think it is
important to understand more about such models which represent bound Universes
(analogous to uni-dimensional atoms, in the present situation).

In the present paper, we use the formalism of quantum cosmology
in order to quantize three Friedmann-Robertson-Walker 
models in the presence of a negative cosmological constant and radiation. 
The radiation is treated by means of the variational formalism developed by Schutz
\cite{schutz}. The models differ from each other by the constant curvature of 
their spatial sections, which may be positive, negative or zero. They give rise 
to Wheeler-DeWitt equations for the scale factor, which have the form of the 
Schr\"{o}dinger equation for the quartic anharmonic oscillator. We find the 
eigenvalues and eigenfunctions of those equations by using a method first 
developed by Chhajlany and Malnev \cite{chhajlany}. Then we use the 
eigenfunctions in order to construct wave packets for each case and evaluate 
the expection value of the scale factors as a function of time.
In Sec. \ref{sec:classical} we introduce the classical
models and solve them analytically, briefly commenting on the general
behavior of the classical solutions. In Sec. \ref{sec:quantization} we 
quantize the model by solving the corresponding Wheeler-DeWitt equation. 
The wave-functional depends on the scale factor 
$a$ and on the canonical variable associated to the fluid, which in the Schutz variational 
formalism plays the role of time $T$. We separate the wave-functional in 
two parts, one depending solely on the scale factor and the other 
depending only on the time. The solution in the time sector of the Wheeler-DeWitt 
equation is trivial, leading to imaginary exponentials of the type 
$e^{-iE\tau}$, where $E$ is the system energy and $\tau =- T$. The scale factor 
sector of the Wheeler-DeWitt equation gives rise to the eigenvalue equation for the quartic 
anharmonic oscillator. We find semi-analytic solutions formed by the product
of a decaying exponential term with a polynomial of fixed degree 
\cite{chhajlany}. In Sec. \ref{sec:results} we
construct wave packets from the eigenfunctions, for each case, and
compute the time-dependent, expectation values of the scale factors. 
We find in all cases that the expection values of the scale factors 
show bounded oscillations. Since the expectation values of the scale factors 
never vanish, we conclude that these models do not have singularities. We also
observe that the energy levels depend on the value of the curvature
constant $k$ of the spatial sections. The model with $k<0$ has the most
bounded energy levels, followed by the one with $k=0$;
the model with $k>0$ have less bounded energy levels.
Finally, in Sec. \ref{sec:conclusions} we summarize the main points and results 
of our paper.
 
\section{The Classical Models}
\label{sec:classical}

Friedmann-Robertson-Walker cosmological models are characterized by the
scale factor $a(t)$ and have the following line element,

\begin{equation}
\label{1}
ds^2 = - N(t)^2 dt^2 + a(t)^2\left( \frac{dr^2}{1 - kr^2} + r^2 d\Omega^2 
\right)\, ,
\end{equation}
where $d\Omega^2$ is the line element of the two-dimensional sphere with unitary radius, 
$N(t)$ is the lapse function and $k$ gives the type of constant curvature
of the spatial sections. The curvature is positive for $k=1$, negative for $k=-1$ and 
zero for $k=0$. Here, we are using the natural unit system,  where 
$\hbar=c=G=1$. The matter content of the model is represented by a 
perfect fluid with four-velocity $U^\mu = \delta^{\mu}_0$ in the comoving 
coordinate system used, plus a negative cosmological constant. The total 
energy-momentum tensor is given by,

\begin{equation}
T_{\mu,\, \nu} = (\rho+p)U_{\mu}U_{\nu} - p g_{\mu,\, \nu} - \Lambda 
g_{\mu,\, \nu}\, ,
\label{2}
\end{equation}
where $\rho$ and $p$ are the energy density and pressure of the fluid, 
respectively. Here, we assume that $p = \rho/3$, which is the equation of 
the state for radiation. This is justified because the Universe is initially dominated
by radiation. Einstein's equations for the metric (\ref{1}) and the
energy momentum tensor (\ref{2}) are equivalent to the Hamilton
equations generated by the super-hamiltonian constraint

\begin{equation}
{\cal{H}}= -\frac{p_{a}^2}{12} - 3ka^2 +\Lambda a^{4} + p_{T},
\label{3}
\end{equation}
where $p_{a}$ and $p_{T}$ are the momenta canonically conjugated to $a$
and $T$ the latter being the canonical variable associated to the fluid \cite{germano1}.

The classical dynamics is governed by the Hamilton equations, derived 
from eq. (\ref{3}), namely

\begin{equation}
\left\{
\begin{array}{llllll}
\dot{a} =&\frac{\partial (\displaystyle N{\cal H})}{\displaystyle
\partial p_{a}}=-\frac{\displaystyle Np_{a}}{\displaystyle 6}\, ,\\
 & \\
\dot{p_{a}} =&-\frac{\displaystyle \partial (N{\cal H})}{\displaystyle 
\partial a}=6kaN-4\Lambda a^3N \, ,\\
& \\
\dot{T} =&\frac{\displaystyle \partial (N{\cal H})}{\displaystyle 
\partial p_{T}}=N\, ,\\
 & \\
\dot{p_{T}} =&-\frac{\displaystyle \partial (N{\cal H})}{\displaystyle 
\partial T}=0\, .\\
& \\
\end{array}
\right.
\label{4}
\end{equation}
We also have the constraint equation ${\cal H} = 0$.

Choosing the gauge $N=1$, we have the following solutions for the
system (\ref{4}):

\begin{equation}
T (\tau) = \tau + c_{1}\, ,\\
a(\tau) = \frac{\sqrt{6\,\beta}}{{{\displaystyle\sqrt{3k+\sqrt {9k^2-12\,\Lambda\,
\beta}}}}} sn\left(\frac{\displaystyle\sqrt{18k+6\,\sqrt{9k^2-12\, \Lambda\,\beta}} 
\left( \tau-\tau_0\right)}{\displaystyle 6},\sigma\right),
\label{5}
\end{equation}
where $c_{1}$, $\beta$ and $\tau_{0}$ are integration constants, $sn$ is the
Jacobi's elliptic sine \cite{abramowitz} of modulus $\sigma$ given by

\begin{equation}
\sigma =\frac{\sqrt{2}}{2}\sqrt{{\frac{-2\beta\Lambda+3k^2-k\sqrt{9k^2-12\,
\Lambda\,\beta}}{\Lambda\,\beta}}}.
\label{6}
\end{equation}
In the case of the models studied here, for wich $\Lambda<0$, Eqs. (\ref{5}) and
(\ref{6})  imply that the scale factor performs bounded oscillations, for all values of $k$. 
When the scale factor vanishes we have the formation
of a singularity which may be either a {\it Big Bang} or a {\it Big Crunch}. For the sake of completeness,
we mention that for $\Lambda=0$, the case studied in Ref. \cite{lemos} is recovered.

\section{The Quantization of the Models}
\label{sec:quantization}

We wish to quantize the models following the Dirac formalism for quantizing 
constrained systems \cite{dirac}. First we introduce a wave-function which is a
function of the canonical variables $\hat{a}$ and $\hat{T}$,

\begin{equation}
\label{7}
\Psi\, =\, \Psi(\hat{a} ,\hat{T} )\, .
\end{equation}
Then, we impose the appropriate commutators between the operators $\hat{a}$ and $\hat{T}$ and
their conjugate momenta $\hat{P}_a$ and $\hat{P}_T$. Working in the Schr\"{o}dinger picture, the operators 
$\hat{a}$ and $\hat{T}$ are simply multiplication operators, while their conjugate momenta are represented by the differential operators
\begin{equation}
p_{a}\rightarrow -i\frac{\partial}{\partial a}\hspace{0.2cm},\hspace{0.2cm}
\hspace{0.2cm}p_{T}\rightarrow -i\frac{\partial}{\partial T}\hspace{0.2cm}.
\label{8}
\end{equation}

Finally, we demand that ${\cal H}$, the super-hamiltonian operator corresponding to (\ref{3}),
annihilate the wave-function $\Psi$, which leads to Wheeler-DeWitt equation
\begin{equation}
\bigg(\frac{1}{12}\frac{{\partial}^2}{\partial a^2} - 3ka^2 + 
\Lambda a^4\bigg)\Psi(a,\tau) = -i \, \frac{\partial}{\partial \tau}\Psi(a,\tau),
\label{9}
\end{equation}
where the new variable $\tau= -T$ has been introduced. In order to avoid possible 
countributions from boundary terms at spatial infinity, we shall consider compact 
tri-dimensional spatial sections in the cases $k=0$ and $k=-1.$

The operator $\hat{{\cal H}}$ is self-adjoint \cite{lemos} with respect to the
internal product,

\begin{equation}
(\Psi ,\Phi ) =  \int_0^{\infty} da\, \,\Psi(a,\tau)^*\, \Phi (a,\tau)\, ,
\label{10}
\end{equation}
if the wave functions are restricted to the set of those satisfying either 
$\Psi (0,\tau )=0$ or $\Psi^{\prime}(0, \tau)=0$, where the prime $\prime$ 
means the partial derivative with respect to $a$. 

The Wheeler-DeWitt equation (\ref{9}) is the Schr\"{o}dinger equation for the
quartic anharmonic oscillator and may be solved by writing $\Psi(a, \tau)$ as
\begin{equation}
\Psi (a,\tau) = e^{-iE\tau}\eta(a)\, 
\label{11}
\end{equation}
where $\eta(a)$ depends solely on $a$. Then $\eta(a)$ satisfies the eigenvalue equation
\begin{equation}
-\frac{d^2{\eta(a)}}{da^2} + V_{e} (a)\eta(a)= 12E\eta(a)\, ,
\label{12}
\end{equation}
where the effective potential $V_{e}(a)$ is given by
\begin{equation}
V_{e}(a) = 36ka^2-12\Lambda a^4\, .
\label{13}
\end{equation}

\subsection{The Method of Chhajlany and Malnev}

The method of Chhajlany and Malnev \cite{chhajlany} starts with the
addition of an extra term to the original anharmonic oscillator
potential, so that the modified Hamiltonian admits a subset of
manifestly normalizable solutions. In the case we are considering, the
extra term to be added to the effective potential (\ref{13}) is
proportional to $a^6$. In terms of that new enlarged potential, the
eigenvalue equation (\ref{12}) may be re-written as 

\begin{equation}
\label{14}
\eta''(a)\, +\, ( \varepsilon - \alpha a^2 - b a^4 - c a^6 ) \eta(a)\, 
=\, 0\,
\end{equation} 
where $\varepsilon = 12E$, $\alpha = 36k$, $b = -12\Lambda$, $c$ is a 
parameter to be determined by the method. The {\it Ansatz} 
for the solution of Eq. (\ref{14}) takes the form
\begin{equation}
\label{15}
\eta(a)\, = N \, \exp{\left(-\frac{c}{4} a^4 - \frac{\gamma}{2} a^2\right)}\, v(a)\, ,
\end{equation}
and has finite norm for $c>0$. Here, $v(a)$ is a polynomial of a certain degree, yet to be chosen; 
the parameter $\gamma$ is to be chosen according to our convenience, as we shall see; $N$ is a normalization
factor. The method is based on the fact, shown in Ref. \cite{chhajlany}, that the larger the degree of the polynomial $v(a)$, the smaller $c$ is. Therefore, if one increases the order of $v(a)$, the energy eigenvalues predicted by the present method tend monotonically, from above, to the energy eigenvalues of the original problem. One important property of the method is that the convergence is very fast. This means that one does not need to use a polynomial
of very large order to obtain a good agreement with the energies of anharmonic oscillators already computed in the literature by other methods.

The next step is the substitution of the {\it Ansatz} (\ref{15}) into the differential equation (\ref{14}), which gives rise to the following equation for the polynomial $v(a)$:
\begin{equation}
\label{16}
v''(a)\, +\, 2( ca^3 + \gamma a ) v'(a)\, +\, [ \varepsilon - \gamma
+ (\gamma^2 + \alpha - 3c)a^2 ] v(a)\, =\, 0\, .
\end{equation}

Next, writing $v(a) = \sum_n \beta_n a^n$ and inserting it in Eq. (\ref{16}) along with the condition  

\begin{equation}
\label{17}
2 c \gamma\, =\, b\, 
\end{equation}
on $c$ and $\gamma$, \cite{chhajlany} we manage to find:

\begin{equation}
\left( \epsilon-{\it \gamma} \right) \beta_{{0}} + 2\,\beta_{{2}} =0,
\qquad
\left( \epsilon-3\,{\it \gamma} \right) \beta_{{1}} + 6\,\beta_{{3}} =0,
\label{18a}
\end{equation}
and the general recurrence relation
for the polynomial coefficients $\beta_n$,

\begin{equation}
\label{18}
(n+4)(n+3)\beta_{n+4}\, +\, [\varepsilon - \gamma (2n+5)]\beta_{n+2}\, 
+\, [\gamma^2 -\alpha - c(2n + 3)]\beta_n\, =\, 0\, ,
\end{equation}
for $n\geq 0$. The degree of the polynomial $v(a)$ is fixed to be, say $K$, by 
imposing the following conditions in (\ref{18}), 

\begin{equation}
\label{19}
\beta_K \neq 0, \qquad \beta_{K+2}\, =\, \beta_{K+4}\, =\, 0\, . 
\end{equation}
Due to the nature of the recurrence relation (\ref{18}), it is
clear that by fixing $K$ to be even (odd) the resulting polynomial
$v(a)$ will be even (odd). Then, the coefficients $\beta_n$, $n=2,4,...,K$
($n=3,5,...,K$), will be determined in terms of $\beta_0$ ($\beta_1$)
by the normalization condition. In the present situation, we restrict 
our attention to the case of an odd polynomial. It means that, 
$K=2m+1$ for $m=0, 1, 2,...$. This condition is imposed in order that 
our wave-function vanishes at $a=0$.

Eqs. (\ref{18}) and (\ref{19}) require that the coeficient $\beta_K$ must vanish; 
then
\begin{equation}
\label{20}
\gamma^2 =\, \alpha\, +\, c(2K + 3)\, .
\end{equation}
Combining this with (\ref{17}), we obtain a  
cubic algebraic equation in the parameter $c$,
\begin{equation}
\label{21}
4 c^3 (2K + 3)\, +\, 4 \alpha c^2\, -\, b^2\, =\, 0\, .
\end{equation}
The solutions of this equation depend on the known parameters $b$ and $K$. 
We must find the real, positive root to this equation so that the {\it Ansatz}
Eq. (\ref{15}) be normalizable. That real positive root,
as proved in Ref. \cite{chhajlany}, is a monotonically
decreasing function of $K$. Therefore, the greater the polynomial
degree, the better the agreement between the energy eigenvalues 
obtained by this method and the actual energy eigenvalues.

Now, by setting the condition $\beta_{K+2}=0$ in Eq. (\ref{18})
we may determine the corresponding energy eigenvalues 
$\varepsilon$ and polynomial coefficients $\beta_n$. The
$(m+1)$ allowed energy levels of the anharmonic oscillator are 
obtained as the roots of the equation $D = 0$, where $D$ is 
the following $(m+1) \times (m+1)$ determinant:
\begin{equation}
\label{22}
%D\, =\, 
\left|\begin{array}{ccccccccc}
(\varepsilon - 3\gamma) & 6 & 0 & 0 & \cdots & & \cdots & & 0 \\
\gamma^2 - \alpha - 5c & (\varepsilon - 7\gamma) & 20 & 0 & \cdots 
& & \cdots & & 0 \\
0 & \gamma^2 - \alpha - 9c & (\varepsilon - 11\gamma) & 42 & \cdots 
& & \cdots & & 0 \\
\cdots & \cdots & \cdots & \cdots & (\gamma^2 - \alpha - (2K+3)c) 
& & \varepsilon - (2K+5)\gamma & & (K+4)(K+3)
\end{array}
\right|.
\end{equation}
The lowest real, positive root will correspond to the ground state
energy level and the excited levels will be given by the sequence
of higher real, positive roots. Next, we must substitute these 
values in the set of Eqs. (\ref{18}) in order to evaluate the
coefficients $\beta_n$ and obtain the appropriate polynomial 
$v_{l}(a)$; the index $l=0,1,2,...,m$ represents the energy level, 
for each of which we shall have an eigenfunction $\eta_{l} (a)$ 
and a wave-function $\Psi_{l} (a, \tau) = exp{(-iE_l\tau)} \eta_{l}(a)$,
according to  Eqs. (\ref{11}) and (\ref{15}).

We construct a general solution to the Wheeler-DeWitt equation (\ref{9})
by taking linear combinations of the $\Psi_{l} (a, \tau)$'s,
\begin{equation}
\Theta (a,\tau) = \sum_{l=0}^{m} A_{l}(E)\eta_l(a)
e^{-iE_{l}\tau},
\label{23}
\end{equation}
where the coefficients $A_{l}(E)$ will be fixed later. 
With those combinations we compute the expected
value for the scale factor $a$, following the {\it many worlds
interpretation} of quantum mechanics \cite{everett}. In the
present situation, we may write the expected value for the scale 
factor $a$ is

\begin{equation}
\left<a\right>(\tau) = \frac{\int_{0}^{\infty}a\,|\Theta (a,\tau)|^2 da}
{\int_{0}^{\infty}|\Theta (a,\tau)|^2 da}.
\label{24}
\end{equation}

\section{Results}
\label{sec:results}

We shall treat, now, each model separately depending on the constant
curvature of the spatial sections. The difference from one model to
the other will appear in the value of the parameter $\alpha$ in Eq.
(\ref{14}). For all models we shall use the value of $\Lambda = -0.1$,
therefore one has $b = 1.2$ in Eq. (\ref{14}). Also, we shall fix the
polynomial degree to be $K = 45$, for all models. It means that, we 
shall have $23$ energy levels and 23 eigenfunctions $\eta_l(a)$. A 
precision of at least 15 significant digits had to be used, in order to guarantee the
orthogonality of the set of functions $\eta_l (a)$'s Eq. (\ref{15}).
The symbolic system Maple has been used, and the precision of
calculations was chosen so that the largest number of energy levels be
achieved and the corresponding (approximate) eigenfunctions be
sufficiently orthogonal.

\subsection{The model with $k=1$.}
\label{subsec:k=1}

In this model $\alpha = 36$ and the spatial sections are $S^3$'s.
Using the values of $K$ and $b$, we solved Eq. (\ref{21}) to find
$c=0.090068960669615962974$. Computations have been performed with
20 significant digits. Now, introducing all these quantities in the
determinant $D$, Eq. (\ref{22}), we obtain the first $23$ energy
levels; they are listed in Table \ref{tableenergy}. The first lowest
energy levels are in agreement with the ones computed, pertubatively,
by Landau for the quartic anharmonic potencial, equivalent to the
present case \cite{landau}. After that, we substitute these values in
the set of Eqs. (\ref{18}) and compute the coefficients $\beta_n$.
With these $\beta_n$, we write the following $\eta (a, \tau)$,
according to Eq. (\ref{15}),

\begin{equation}
\eta_l(a)= {N}_{l} \ {e^{- 0.022517240167403990744\,{a}^{4}- 
3.3307811899866029985\,{a}^{2}}}\ v_l(a),
\label{25}
\end{equation}
where
\begin{equation}
\label{26}
v_l(a)=\sum_{i= 0}^{22} A_{l,2i+1} \ a^{2i+1}.
\end{equation}
The coefficients ${N}_{l}$ are normalization coefficients and 
$i=0, 1,..., m$.
The $N_{l}$'s and the $A_{l,2i+1}$'s for the present model are
listed in the appendix \ref{A1}, Tables \ref{tabela2} to
\ref{tabela7}.

\begin{figure}[h!]
\includegraphics[width=6cm,height=9cm,angle=-90]{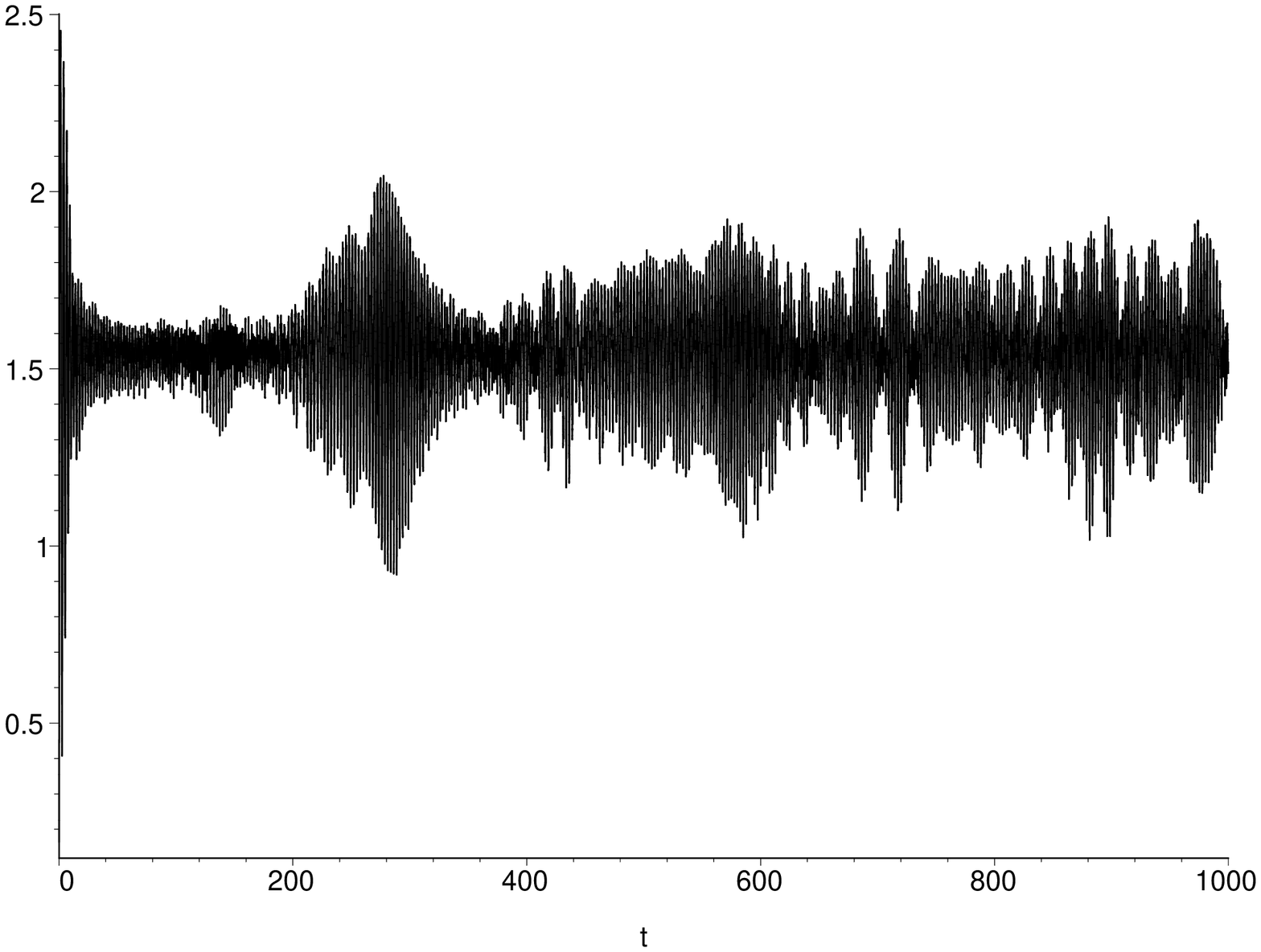} 
\mycaption{Behavior of the expectation value of the scalar factor for $\Lambda=-0.1$, $k=1$.}
\label{f1}
\end{figure}

Next, we construct the wave-packet $\Theta (a, \tau )$ with the aid of
the $\eta_l (a)$, according to Eqs. (\ref{23}), Eq. (\ref{25}), and
the energy levels in Table \ref{tableenergy}. Finally, using the
wave-packet $\Theta (a, \tau )$ we compute the expected value for the
scale factor $a$, Eq. (\ref{24}). The result is shown in Fig.
\ref{f1}; it can be seen that $\left<a\right>$ does not vanish, therefore we may
say that the quantization of this model removed the singularities it
had at the classical level. It is clear from Fig. \ref{f1}, also,
that $\left<a\right>$ performs bounded oscillations. 
That means that the spatial sections $S^3$'s oscillate between
finite maximum and minimum radius.

\subsection{The model with $k=0$}
\label{subsec:k=0}

In this model  $\alpha = 0$ and the spatial sections
are some closed three-dimensional solid with zero curvature, locally 
isometric to $R^3$ \cite{wolf}. Here, like in the previous case, we have used 20 
significant digits. Introducing the values of $K$ and $b$ in Eq. 
(\ref{21}) we obtain $c=0.15701453260387612225$. Now, using all these 
quantities in the determinant $D$ Eq. (\ref{22}), we obtain
the first $23$ energy levels. They are shown in Table 
\ref{tableenergy}. The first lowest energy levels are in agreement 
with the ones computed, pertubatively, by Landau for the quartic 
anharmonic potencial, equivalent to the present case 
\cite{landau}. After that, we substitute these values in the 
set of Eqs. (\ref{18}) and compute the coefficients $\beta_n$. 
With these $\beta_n$, we write the following $\eta (a, \tau)$ 
Eq. (\ref{15}),

\begin{equation}
\eta_l(a)=N_{l} \ {e^{- 0.039253633150969030562\,{a}^{4}- 
1.9106511672830600300\,{a}^{2}}}\ v_l(a),
\label{27}
\end{equation}
where the $v_{l}(a)$ have the general expression given in 
Eq. (\ref{26}) and the coefficients $N_{l}$ are normalization 
coefficients. The $N_{l}$'s and the $A_{l,2i+1}$'s for the 
present model are listed in appendix \ref{A1}, Tables \ref{tabela8} 
to \ref{tabela13}.

\begin{figure}[h!]
\includegraphics[width=6cm,height=9cm,angle=-90]{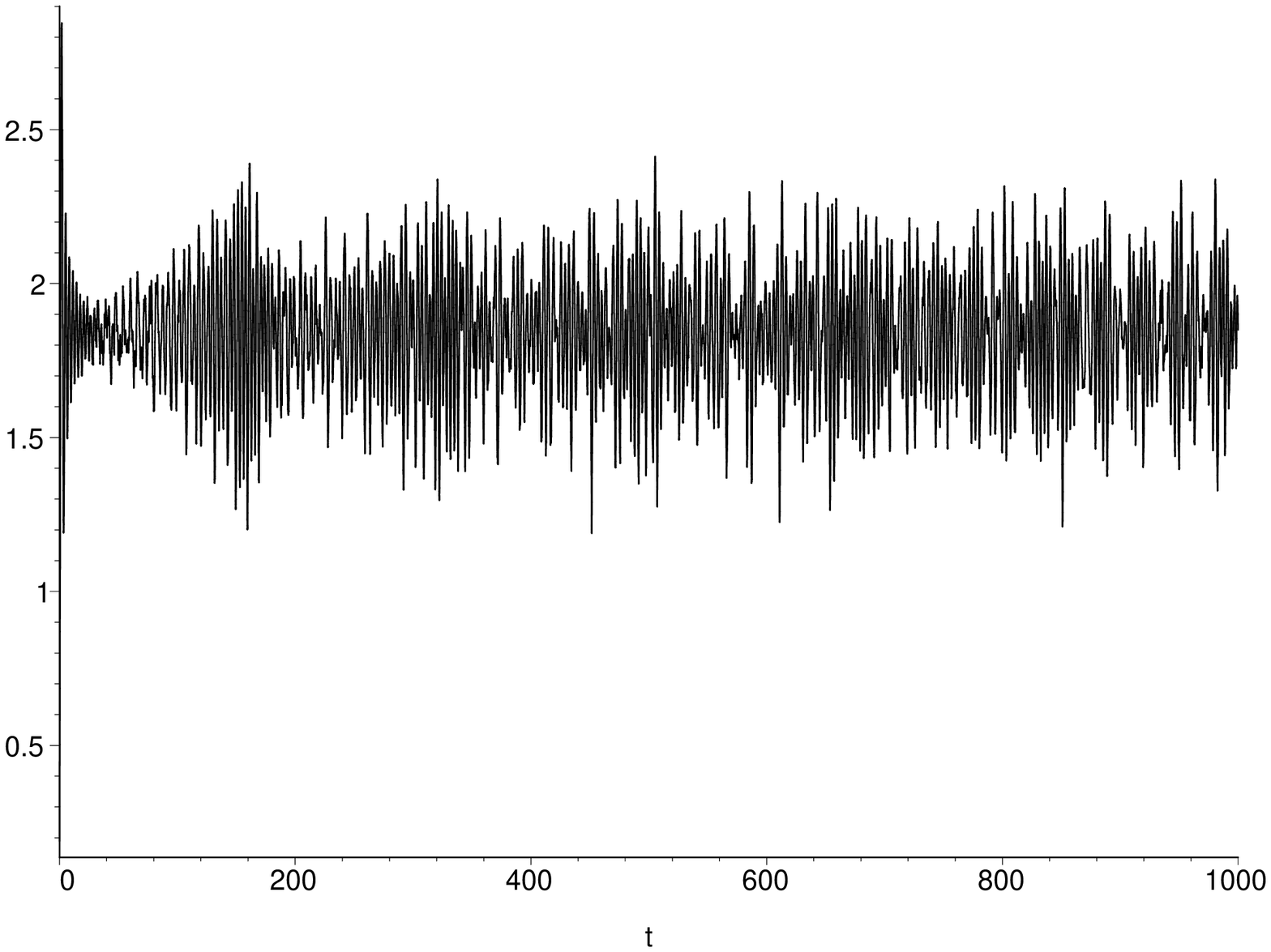}
\mycaption{Behavior of the expectation value of the scalar factor for $\Lambda=-0.1$, $k=0$.}
\label{f2}
\end{figure}

Next, we construct the wave-packet $\Theta (a, \tau )$, with the aid
of the $\eta_l (a)$ and the energy levels, according to Eqs.
(\ref{23}) and (\ref{27}), as well as Table \ref{tableenergy}. Using
the wave-packet $\Theta (a, \tau )$ we compute the expected value for
the scale factor $a$, as in Eq. (\ref{24}). The result is shown in
Fig. \ref{f2}. It can be seen that $\left<a\right>$ does not assume
the value zero; therefore we may say that the quantization of this
model removed the singularities it had at the classical level. It is
clear, also, from Fig. \ref{f2}, that $\left<a\right>$ has bounded
oscillations, that is, oscillates between finite maximum and minimum
values. 

\subsection{The model with $k=-1$}
\label{subsec:k=-1}

In this model $\alpha = - 36$ and the spatial sections are
some closed three-dimensional solid with negative constant curvature,
locally isometric to $H^3$ \cite{thurston}. Here, we have used 15 
significant digits. Introducing the values of $K$ and $b$, we 
solved Eq. (\ref{21}) to
find $c = 0.410111969406177$. Now, using all these quantities in the
determinant $D$ Eq. (\ref{22}), we obtain the first $23$ energy
levels; they are listed in Table \ref{tableenergy}. After that, we
substitute these values in the set of Eqs. (\ref{18}) and compute the
coefficients $\beta_n$, used in 

\begin{equation}
\eta_l(a)=N_{l} \ {e^{- 0.102527992350000\,{a}^{4}- 
0.731507545000000\,{a}^{2}}}\ v_l(a),
\label{29}
\end{equation}
where the $v_{l}(a)$ have the general expression given in Eq. (\ref{26})
and the $N_{l}$'s are normalization coefficients.
The $N_{l}$'s and the $A_{l,2i+1}$'s for the present model are listed 
in the appendix \ref{A1}, Tables \ref{tabela14} to \ref{tabela18}. Due
to numerical inconsistencies we have considered, in the present case,
$18$ $v_{l}(a)$'s corresponding to the first $18$ energy levels.

\begin{figure}[h!]
\includegraphics[width=6cm,height=9cm,angle=-90]{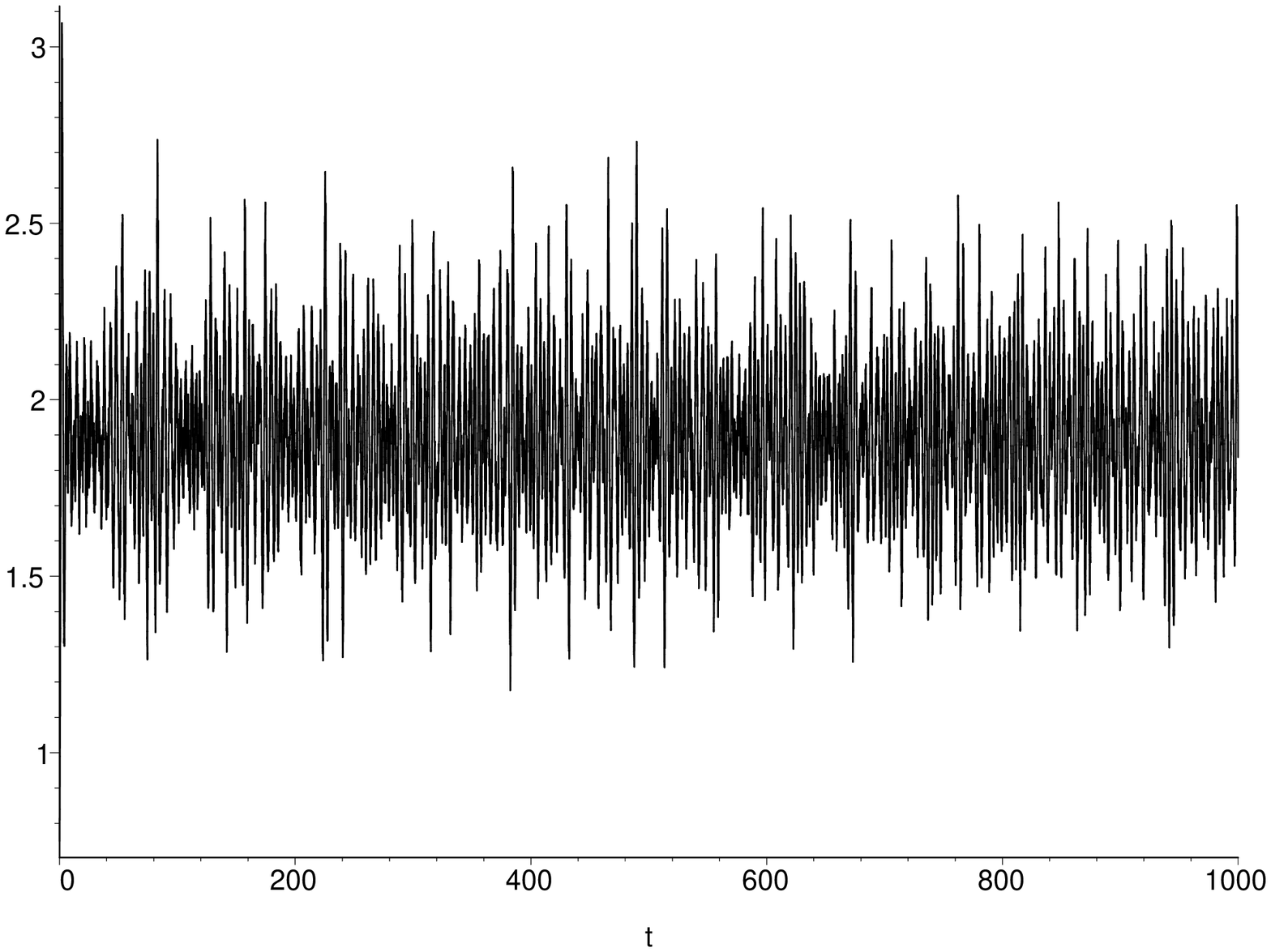}
\mycaption{Behavior of the expectation value of the scalar factor for $\Lambda=-0.1$, $k=-1$.}
\label{f3}
\end{figure}

Next, we construct the wave-packet $\Theta (a, \tau )$ with the aid of
the $\eta_l (a)$, according to Eqs. (\ref{23}) and (\ref{29}) and the
energy levels in Table \ref{tableenergy}. Finally, using the
wave-packet $\Theta (a, \tau )$ we compute the expected value for the
scale factor $a$, according to (\ref{24}). The result is shown in Fig.
\ref{f3}. It can be seen that $\left<a\right>$ does not assume the
value zero. Therefore, we may say that the quantization of this model
removed the singularities it had at the classical level. It is also
clear that $\left<a\right>$ performs bounded oscillations.

From Table \ref{tableenergy}, we observe that the energy levels 
depend on the value of the  curvature constant of the spatial 
sections. The model with negative constant curvature has the 
most bounded energy levels, then one has the model with zero 
curvature and finally the model with positive constant 
curvature has the less bounded energy levels.

Observing Eqs. (\ref{25}), (\ref{27}) and (\ref{29}), we see that the
wave-functions $\Psi (a, \tau)$ for all three cases are exponentially
damped as $a\rightarrow\infty$ and behave as powers of $a$ in the
limit $a\rightarrow 0$. Following Hawking and Page \cite{hawking1}, we
may say that this behavior of the $\Psi (a, \tau)$'s makes
them {\it excited states} of wormholes.

\begin{table}[h!]
{\scriptsize\begin{tabular}{|c|c|c|c|}
\hline Level & $k=1$ & $k=0$  &  $k=-1$\\ \hline
$E_{1}$ & $1.5103016760578712464$ & $0.34040279742876372340$ & $-10.5531383749465$ \\ \hline
$E_{2}$ & $3.5509871014722954423$  & $1.0496018816133576012$  & $-8.85096087898748$  \\ \hline
$E_{3}$ &  $5.6230931685080087038$ & $1.9248782225962522139$  & $-7.22478356024350$ \\ \hline
$E_{4}$ & $7.7256531719671366439$ & $2.9232145294773461677$  & $-5.68029499265656$  \\ \hline
$E_{5}$ & $9.8577817762723638293$  & $4.0227383298683634726$  & $-4.22519889374100$ \\ \hline 
$E_{6}$ & $12.018664367453930157$  & $5.2097960735548422725$  & $-2.87075159719689$ \\ \hline
$E_{7}$ & $14.207548216681935132$  & $6.4748734067709572007$  & $-1.63482529734928$ \\ \hline
$E_{8}$ & $16.423735430300010131$ & $7.8108774877366064367$  & $-0.536430728059501$ \\ \hline
$E_{9}$ & $18.666574187417793427$  & $9.2122719218977134548$  &  $0.481759230982801$ \\ \hline
$E_{10}$  & $20.935469528987599984$  & $10.674588231181872438$  & $1.58319085808971$ \\ \hline
$E_{11}$  & $23.229800589269486517$  & $12.194126672497696030$  & $2.83303176055370$ \\ \hline
$E_{12}$  & $25.549220854858484858$  & $13.767762525648273494$  & $4.21398303307806$ \\ \hline
$E_{13}$  & $27.892697492531834013$  & $15.392811058629387358$  & $5.70783697327336$ \\ \hline
$E_{14}$  & $30.260978882469751228$  & $17.066942189323770737$  & $7.30286131347368$ \\ \hline
$E_{15}$  & $32.651319599290796010$ & $18.788091721482546958$  & $8.99091730871327$ \\ \hline
$E_{16}$  & $35.066840708164574204$  & $20.554451348143454694$  & $10.7657478496350$ \\ \hline
$E_{17}$ & $37.502643477205645295$  & $22.364363322148552407$   & $12.6225505157144$ \\ \hline
$E_{18}$ & $39.963027226267456788$  & $24.216376790909429384$  & $14.5570952852202$ \\ \hline
$E_{19}$ & $42.443644402959303694$  & $26.109132260627772146$  & $16.5661310939634$ \\ \hline
$E_{20}$ & $44.946892054208187300$  & $28.041419503470901850$  & $18.6464752057244$ \\ \hline
$E_{21}$ & $47.470929938318337116$  & $30.012110489502592151$  & $20.7956557497686$ \\ \hline
$E_{22}$ & $50.01608754545613080 $ & $32.020172678051097447$  & $23.0112449804346$ \\ \hline
$E_{23}$ & $52.581852999800707709$ & $34.064648529603099090$  & $25.2911792258758$ \\ \hline
\end{tabular}}
\mycaption{The lowest calculated energy levels for the cases $k=0$, $k=1$, and $k=-1$ (in all cases,
$\Lambda=-0.1$).}
\label{tableenergy}
\end{table}

\section{Conclusions.}
\label{sec:conclusions}

In the present paper, the formalism of quantum cosmology was employed to quantize three Friedmann-Robertson-Walker
models in the presence of a negative cosmological constant and
radiation. The variational formalism of Schutz \cite{schutz} allowed us to ascribe dynamical 
degrees of freedom to the radiation fluid. The models differ from each other
by the constant curvature of the spatial sections, which may be
positive, negative or zero. The quantization of the models gave rise to Wheeler-DeWitt
equations, for the scale factor, which had the form of the
Schr\"{o}dinger equation for the quartic anharmonic-oscillator. We
found the approximate eigenvalues and eigenfunctions of those equations by using a
method first developed by Chhajlany and Malnev \cite{chhajlany}. After
that, we used the eigenfunctions in order to construct wave-packets
for each case and evaluate the time dependent, expected value of the
scale factors. We found for all of them that the expected values of
the scale factors evolve with bounded oscillations. Since the 
expectation value of the scale
factors never vanish, we concluded that these models do not have
singularities. We also observed that the energy levels depend on the
value of the curvature constant of the spatial sections. The model
with negative curvature constant has the most bounded energy levels,
whereas the model with positive constant curvature has the less 
bounded energy levels.

\begin{acknowledgements}
E. V. Corr\^{e}a Silva thanks CNPq for partial financial support.
\end{acknowledgements}

\appendix
\section{Coefficients of the Polynomials $v_i(a)$}\label{A1}

The following Tables contain the coefficients $A_{l,2i+1}$ introduced 
in eq. (\ref{26}), for each
case considered. Tables \ref{tabela2} to \ref{tabela7} refer to the
case $\Lambda=-0.1$, $k=1$; Tables \ref{tabela8} to \ref{tabela13}
refer to the case $\Lambda=-0.1$, $k=0$; Tables \ref{tabela14} to
\ref{tabela18} refer to the case $\Lambda=-0.1$, $k=-1$. The
normalization coefficients $N_l$ can be found in the respective
captions.

\begin{table}
{\tiny
\begin{tabular}{|l|l|l|l|l|} 
\hline 
$i$ & Coefficients $A_{1,2i+1}$  & Coefficients $A_{2,2i+1}$  & Coefficients $A_{3,2i+1}$ & 
Coefficients $A_{4,2i+1}$ \\ \hline 
$0$ & $1$ & $1$ & $1$ & $1$\\ \hline 
$1$ & $
 0.31017783787086050567$ & $- 3.7711930129579878862$ & $-
 7.9154051470294144092$ & $- 12.120525153947670289$\\ \hline 
$2$ & $
 0.04581346355795285829$ & $- 1.1541419051991141135$ & $ 7.853995135600127347$ & $
 27.527508697085798433$\\ \hline 
$3$ & $ 0.004286539733354114758$ & $-
 0.1633354878078236419$ & $ 2.510476703458296824$ & $- 10.551810905706899599
$\\ \hline 
$4$ & $ 0.0002851177209318121384$ & $-
 0.014511494987299668478$ & $ 0.3453281989577501333$ & $- 3.812327078892842085
$\\ \hline 
$5$ & $ 0.000014340884997309905358$ & $-
 0.0009116635877726964358$ & $ 0.02928647013334059913$ & $-
 0.5169340736618381480$\\ \hline 
$6$ & $ 0.000000566441403839108$ & $-
 0.0000431229586177813$ & $ 0.001740679187283938$ & $- 0.042024926844112081
$\\
\hline
\end{tabular}}
\mycaption{The case $\Lambda=-0.1$, $k=1$: coefficients $A_{1,2i+1}$,
	$A_{2,2i+1}$, $A_{3,2i+1}$ and $A_{4,2i+1}$ of the polinomyals
	$v_{1}$, $v_{2}$, $v_{3}$ and $v_{4}$. The coefficients of
	the normalization are $N_{1}=5.8034051393293824531$,
	$N_{2}=7.1778933987585237889$, $N_{3}=8.1009761694435305313$ and
	$N_{4}=8.8293734529556238792$.}
\label{tabela2}
\end{table}

\begin{table}
{\tiny\begin{tabular}{|l|l|l|l|l|}\hline 
 $ i$ & Coefficients $A_{5,2i+1}$ & Coefficients $A_{6,2i+1}$ & Coefficients $A_{7,2i+1}$ & Coefficients $A_{8,2i+1}$ \\ \hline 
 $0$ & $1$ & $1$ & $1$ & $1$ \\ \hline 
 $1$ & $-
 16.384782362558124660$ & $- 20.706547544921257315$ & $- 25.084315243377267265
$ & $- 29.516689670613417263$ \\ \hline 
 $2$ & $ 58.312374535479288149$ & $
 100.64443831320425508$ & $ 154.95041179125834137$ & $ 221.64873340822231830
$ \\ \hline 
 $3$ & $- 59.54850867092679165$ & $- 166.27947628665417837$ & $-
 354.13029662189362099$ & $- 648.05926695036679593$ \\ \hline 
 $4$ & $
 9.357389837391042752$ & $ 92.23715065066104101$ & $ 331.57590893769938406$ & $
 852.35062705243081723$ \\ \hline 
 $5$ & $ 4.409713597601337702$ & $-
 4.45598233475616519$ & $- 110.35457194930656034$ & $- 506.0681988766809666
$ \\ \hline 
 $6$ & $ 0.59818020215651014$ & $- 4.09117137488257791$ & $-
 1.56360443105183866$ & $ 106.8793989497608568$ \\
\hline \end{tabular}}
\mycaption{The case $\Lambda=-0.1$, $k=1$: coefficients $A_{5,2i+1}$,
	$A_{6,2i+1}$, $A_{7,2i+1}$ and $A_{8,2i+1}$ of the polinomyals
	$v_{5}$, $v_{6}$, $v_{7}$ and $v_{8}$. The coefficients of the
	normalization are $N_{5}=9.4465594769581691233$,
	$N_{6}=9.9907608238646431121$, $N_{7}=10.482871871428757467$ and
	$N_{8}=10.935645654327260289$.}
\label{table3}
\end{table}

\begin{table}
{\tiny\begin{tabular}{|l|l|l|l|l|}\hline $i$ & Coefficients $A_{9,2i+1}$ & Coefficients $A_{10,2i+1}$ & Coefficients $A_{11,2i+1}$ & Coefficients $A_{12,2i+1}$ \\ \hline 
 $0$ & $1$ & $1$ & $1$ & $1$ \\ \hline 
 $1$ & $-
 34.002367184848983855$ & $- 38.540157867988596970$ & $- 43.128819988552370035
$ & $- 47.767660519730366718$ \\ \hline 
 $2$ & $ 301.15021081131710573$ & $
 393.85929396557445690$ & $ 500.17116570886464547$ & $ 620.48706397200240588
$ \\ \hline 
 $3$ & $- 1074.5861450716873336$ & $- 1661.7872217348979949$ & $
- 2439.2637603904004098$ & $- 3438.2520498616334939$ \\ \hline 
 $4$ & $
 1821.6641872613443414$ & $ 3452.6958410202546019$ & $ 6008.6072441957345911$ & $
 9806.9867235942380195$ \\ \hline 
 $5$ & $- 1546.6135676876763965$ & $-
 3809.3067552029814704$ & $- 8161.2736489861159871$ & $- 15835.705001515897835
$ \\ \hline 
 $6$ & $ 626.0285291644673913$ & $ 2249.7333239006545920$ & $
 6318.0405997660246921$ & $ 15161.530656034483683$ \\
\hline \end{tabular}}  
\mycaption{The case $\Lambda=-0.1$, $k=1$: coefficients $A_{9,2i+1}$,
	$A_{10,2i+1}$, $A_{11,2i+1}$ and $A_{12,2i+1}$ of the polinomyals
	$v_{9}$, $v_{10}$, $v_{11}$ and $v_{12}$. The coefficients of the
	normalization are $N_{9}=11.357473916077508835$,
	$N_{10}=11.754220307373787325$, $N_{11}=12.130088843440861937$ and
	$N_{12}=12.488270658573768104$.}
\label{tabela4}
\end{table}

\begin{table}
{\tiny\begin{tabular}{|l|l|l|l|l|}\hline $i$ & Coefficients $A_{13,2i+1}$ & Coefficients $A_{14,2i+1}$ & Coefficients $A_{15,2i+1}$ & Coefficients $A_{16,2i+1}$ \\ \hline 
 $0$ & $1$ & $1$ & $1$ & $1$ \\ \hline 
 $1$ & $-
 52.454613795077065028$ & $- 57.191176574952899458$ & $- 61.971858008594989022
$ & $- 66.802900226342545410$ \\ \hline 
 $2$ & $ 755.16371270759864284$ & $
 904.65638191366384671$ & $ 1069.1911724988718290$ & $ 1249.3896027640753475
$ \\ \hline 
 $3$ & $- 4691.1786486789463820$ & $- 6233.0019360858040813$ & $
- 8097.8534617243312174$ & $- 10325.918637224436785$ \\ \hline 
 $4$ & $
 15222.159586897177725$ & $ 22695.269368957207490$ & $ 32722.191703657035811$ & $
 45893.949491372741828$ \\ \hline 
 $5$ & $- 28511.535588496471157$ & $-
 48419.757316339959116$ & $- 78400.032812948401888$ & $- 122116.39684597553325
$ \\ \hline 
 $6$ & $ 32538.633947802319935$ & $ 64210.863046864482262$ & $
 118551.52404606112225$ & $ 207557.92486531471539$ \\
\hline \end{tabular}}  
\mycaption{The case $\Lambda=-0.1$, $k=1$: coefficients $A_{13,2i+1}$,
	$A_{14,2i+1}$, $A_{15,2i+1}$ and $A_{16,2i+1}$ of the polinomyals
	$v_{13}$, $v_{14}$, $v_{15}$ and $v_{16}$. The coefficients of the
	normalization are $N_{13}=12.831657346516500646$,
	$N_{14}=13.160199699505029703$, $N_{15}=13.481183098459831301$ and
	$N_{16}=13.783901938149622931$.}
\label{tabela5}
\end{table}

\begin{table}
{\tiny\begin{tabular}{|l|l|l|l|l|}\hline $ i$ & Coeficientes $A_{17,2i+1}$ & Coeficientes $A_{18,2i+1}$ & Coeficientes $A_{19,2i+1}$ & Coeficientes $A_{20,2i+1}$ \\ \hline 
 $0$ & $1$ & $1$ & $1$ & $1$ \\ \hline 
 $1$ & $-
 71.674505764424687592$ & $- 76.595273262548310578$ & $- 81.556507615932004390
$ & $- 86.563002918429771602$ \\ \hline 
 $2$ & $ 1445.2812913047251820$ & $
 1657.6056242524762461$ & $ 1886.3841243300680194$ & $ 2132.2207701020935152
$ \\ \hline 
 $3$ & $- 12951.762075898028761$ & $- 16020.740104978502468$ & $
- 19569.862939305032690$ & $- 23646.245366475767022$ \\ \hline 
 $4$ & $
 62834.794759969799382$ & $ 84306.333013897444312$ & $ 111087.71291195023833$ & $
 144107.29200810027984$ \\ \hline 
 $5$ & $- 183963.75221484100112$ & $-
 269539.48654730779802$ & $- 385321.01208259817624$ & $- 539314.72674103090793
$ \\ \hline 
 $6$ & $ 347409.38740624855152$ & $ 560351.91063946410830$ & $
 874970.62323112283525$ & $ 1328975.0572268366759$ \\
\hline \end{tabular}}  
\mycaption{The case $\Lambda=-0.1$, $k=1$: coefficients $A_{17,2i+1}$,
	$A_{18,2i+1}$, $A_{19,2i+1}$ and $A_{20,2i+1}$ of the polinomyals
	$v_{17}$, $v_{18}$, $v_{19}$ and $v_{20}$. The coefficients of the
	normalization are $N_{17}=14.088838258318067966$,
	$N_{18}=14.371929657653422158$, $N_{19}=14.659291915176228163$ and
	$N_{20}=14.930156447596541712$.}
\label{tabela6}
\end{table}

\begin{table}
{\tiny\begin{tabular}{|l|l|l|l|}\hline $i$ & Coefficients $A_{21,2i+1}$ & Coefficients $A_{22,2i+1}$ & Coefficients $A_{23,2i+1}$ \\ \hline 
 $0$ & $1$ & $1$ & $1$ \\ \hline 
 $1$ & $-
 71.674505764424687592$ & $- 76.595273262548310578$ & $- 81.556507615932004390
$ \\ \hline 
 $2$ & $ 1445.2812913047251820$ & $ 1657.6056242524762461$ & $
 1886.3841243300680194$ \\ \hline 
 $3$ & $- 12951.762075898028761$ & $-
 16020.740104978502468$ & $- 19569.862939305032690$ \\ \hline 
 $4$ & $
 62834.794759969799382$ & $ 84306.333013897444312$ & $ 111087.71291195023833
$ \\ \hline 
 $5$ & $- 183963.75221484100112$ & $- 269539.48654730779802$ & $
- 385321.01208259817624$ \\ \hline 
 $6$ & $ 347409.38740624855152$ & $
 560351.91063946410830$ & $ 874970.62323112283525$ \\
\hline \end{tabular}}  
\mycaption{The case $\Lambda=-0.1$, $k=1$: coefficients $A_{21,2i+1}$,
	$A_{22,2i+1}$ and $A_{23,2i+1}$ of the polinomyals $v_{21}$, $v_{22}$
	and $v_{23}$. The coefficients of the normalization are
	$N_{21}=15.196350124389910569$, $N_{22}=15.443515484903478896$ and
	$N_{23}=15.645567647056785661$.}
\label{tabela7}
\end{table}

\begin{table}
{\tiny\begin{tabular}{|l|l|l|l|l|}\hline $ i$ & Coefficients $A_{1,2i+1}$ & Coefficients $A_{2,2i+1}$ & Coefficients $A_{3,2i+1}$ & Coefficients $A_{4,2i+1}$ \\ \hline 
 $0$ & $1$ & $1$ & $1$ & $1$ \\ \hline 
 $1$ & $
 1.2298455724255325832$ & $- 0.18855259594365517233$ & $-
 1.9391052779094443978$ & $- 3.9357778916716323053$ \\ \hline 
 $2$ & $
 0.70281444756044962535$ & $- 0.8243016140395259213$ & $-
 1.0448066649011448432$ & $ 0.9482808911585986722$ \\ \hline 
 $3$ & $
 0.24882730009488482050$ & $- 0.51856758078952620652$ & $
 0.1378821684225774495$ & $ 1.3929964156767481609$ \\ \hline 
 $4$ & $
 0.061362228801690876509$ & $- 0.17831099849711656507$ & $
 0.24781182769474065702$ & $ 0.26486205576426314590$ \\ \hline 
 $5$ & $
 0.011229566129419549313$ & $- 0.041020332973195298495$ & $
 0.096571463561412127319$ & $- 0.06075921121492778053$ \\ \hline 
 $6
$ & $ 0.0015858480631393467966$ & $- 0.006876928456515549490$ & $
 0.022150497291839980024$ & $- 0.039763176718287530660$ \\
\hline \end{tabular}}  
\mycaption{The case $\Lambda=-0.1$, $k=0$: coefficients $A_{1,2i+1}$,
	$A_{2,2i+1}$, $A_{3,2i+1}$ and $A_{4,2i+1}$ of the polinomyals
 $v_{1}$, $v_{2}$, $v_{3}$ and $v_{4}$. The coefficients of the
 normalization are
	$N_{1}=2.1407112737145309126$, $N_{2}=3.2975672670529412541$,
	$N_{3}=4.1572580515879983879$ and $N_{4}=4.8776996852430886717$.}
\label{tabela8}
\end{table}

\begin{table}
{\tiny\begin{tabular}{|l|l|l|l|l|}\hline $ i$ & Coefficients $A_{5,2i+1}$ & Coefficients $A_{6,2i+1}$ & Coefficients $A_{7,2i+1}$ & Coefficients $A_{8,2i+1}$ \\ \hline 
 $0$ & $1$ & $1$ & $1$ & $1$ \\ \hline 
 $1$ & $-
 6.1348254924536669152$ & $- 8.5089409798266245150$ & $- 11.039095646258854371
$ & $- 13.711103808190152843$ \\ \hline 
 $2$ & $ 5.911356608106969186$ & $
 14.526711829815446250$ & $ 27.430881469808772742$ & $ 45.228591749889808489
$ \\ \hline 
 $3$ & $ 1.048461113886525862$ & $- 4.412562811395891946$ & $-
 19.826241542993934663$ & $- 51.364469972481314193$ \\ \hline 
 $4$ & $-
 0.8995614524956676092$ & $- 2.215775126916415019$ & $ 0.826029728265588547$ & $
 18.084829935024654405$ \\ \hline 
 $5$ & $- 0.31272201167583072358$ & $
 0.2754965915900852529$ & $ 2.112554853993957234$ & $ 2.098921351752975453
$ \\ \hline 
 $6$ & $- 0.01422790600763145292$ & $ 0.20538109262592843443
$ & $ 0.07815218589511349951$ & $- 1.3891586577302272466$ \\
\hline \end{tabular}}  
\mycaption{The case $\Lambda=-0.1$, $k=0$: coefficients $A_{5,2i+1}$,
	$A_{6,2i+1}$, $A_{7,2i+1}$ and $A_{8,2i+1}$ of the polinomyals
	$v_{5}$, $v_{6}$ $v_{7}$ and $v_{8}$. The coefficients of the
	normalization are $N_{5}=5.5127641866644550323$,
	$N_{6}=6.0885073250502348952$, $N_{7}=6.6199175722251377252$ and
	$N_{8}=7.1165584094285250131$.}
\label{tabela9}
\end{table}

\begin{table}
{\tiny\begin{tabular}{|l|l|l|l|l|}\hline $ i$ & Coefficients $A_{9,2i+1}$ & Coefficients $A_{10,2i+1}$ & Coefficients $A_{11,2i+1}$ & Coefficients $A_{12,2i+1}$ \\ \hline 
 $0$ & $1$ & $1$ & $1$ & $1$ \\ \hline 
 $1$ & $-
 16.513892676512366880$ & $- 19.438525295080684847$ & $- 22.477602177712332030
$ & $- 25.624873884013486958$ \\ \hline 
 $2$ & $ 68.500816128447368264$ & $
 97.809919332667553061$ & $ 133.70317321896426872$ & $ 176.71530644883325203
$ \\ \hline 
 $3$ & $- 106.55682738798604833$ & $- 194.31462763518727966$ & $
- 324.95562820472674522$ & $- 510.22884819063511319$ \\ \hline 
 $4$ & $
 66.82399915663488053$ & $ 173.94601726151463769$ & $ 378.40039317842692369$ & $
 733.75951884021536989$ \\ \hline 
 $5$ & $- 11.49015530870562177$ & $-
 66.66856760315107297$ & $- 218.36140595295844410$ & $- 562.39713852806645723
$ \\ \hline 
 $6$ & $- 3.173797462030778022$ & $ 4.57659601834016473$ & $
 54.37892239826436592$ & $ 225.58435351580697565$ \\
\hline \end{tabular}}  
\mycaption{The case $\Lambda=-0.1$, $k=0$: coefficients $A_{9,2i+1}$,
	$A_{10,2i+1}$, $A_{11,2i+1}$ and $A_{12,2i+1}$ of the polinomyals
	$v_{9}$, $v_{10}$ $v_{11}$ and $v_{12}$. The coefficients of the
	normalization are $N_{9}=7.5849746655368418037 $,
	$N_{10}=8.0298788484656264819$, $N_{11}=8.4547979707038394024$ and
	$N_{12}=8.8624524914410403455$.}
\label{tabela10}
\end{table}

\begin{table}
{\tiny\begin{tabular}{|l|l|l|l|l|}\hline $ i$ & Coefficients $A_{13,2i+1}$
 & Coefficients $A_{14,2i+1}$ & Coefficients $A_{15,2i+1}$ & Coefficients $A_{16,2i+1}$ \\ \hline 
 $0$ & $1$ & $1$ & $1$ & $1$ \\ \hline 
 $1$ & $-
 28.874970949975714687$ & $- 32.223233211364481443$ & $- 35.665532275682033887
$ & $- 39.198251529003849358$ \\ \hline 
 $2$ & $ 227.37032148479115396$ & $
 286.18322037516669316$ & $ 353.66043745913659654$ & $ 430.30233892012025033
$ \\ \hline 
 $3$ & $- 763.33867329688880573$ & $- 1098.9702787419893209$ & $
- 1533.3090062044208992$ & $- 2084.0733056378894925$ \\ \hline 
 $4$ & $
 1310.9566981411351041$ & $ 2201.1873227411010990$ & $ 3518.9507974024094019$ & $
 5405.2859712063233703$ \\ \hline 
 $5$ & $- 1253.2820823144540218$ & $-
 2526.1788310714302717$ & $- 4723.4798592947973963$ & $- 8326.4852691380756713
$ \\ \hline 
 $6$ & $ 682.8659047522031438$ & $ 1733.7148182448332023$ & $
 3910.3445264112686706$ & $ 8082.2732257288263686$ \\
\hline \end{tabular}}  
\mycaption{The case $\Lambda=-0.1$, $k=0$: coefficients $A_{13,2i+1}$,
	$A_{14,2i+1}$, $A_{15,2i+1}$ and $A_{16,2i+1}$ of the polinomyals
	$v_{13}$, $v_{14}$ $v_{15}$ and $v_{16}$. The coefficients of the
	normalization are $N_{13}=9.2549982289601205306$,
	$N_{14}=9.6341574144155898259$, $N_{15}=10.001394801227172017$ and
	$N_{16}=10.357816917312201220$.}
\label{tabela11}
\end{table}

\begin{table}
{\tiny\begin{tabular}{|l|l|l|l|l|}\hline $i$ & Coefficients $A_{17,2i+1}$ & Coefficients $A_{18,2i+1}$ & Coefficients $A_{19,2i+1}$ & Coefficients $A_{20,2i+1}$ \\ \hline 
 $0$ & $1$ & $1$ & $1$ & $1$ \\ \hline 
 $1$ & $-
 42.818075477014044783$ & $- 46.522102414535798738$ & $- 50.307613353972484262
$ & $- 54.172187839658743670$ \\ \hline 
 $2$ & $ 516.60124996673597985$ & $
 613.04593626399226781$ & $ 720.11780439718317891$ & $ 838.29525508507567610
$ \\ \hline 
 $3$ & $- 2770.5184363380461860$ & $- 3613.4869945862085907$ & $
- 4635.3958915081377870$ & $- 5860.2905303705903531$ \\ \hline 
 $4$ & $
 8031.0643052243792793$ & $ 11600.594264758079964$ & $ 16355.134985764342392$ & $
 22576.823483403682361$ \\ \hline 
 $5$ & $- 13992.376573406155088$ & $-
 22597.400379554647521$ & $- 35285.868470667908997$ & $- 53526.528942521508014
$ \\ \hline 
 $6$ & $ 15606.322037343566342$ & $ 28522.411642485439145$ & $
 49802.769083156847083$ & $ 83665.654138659815313$ \\
\hline \end{tabular}}  
\mycaption{The case $\Lambda=-0.1$, $k=0$: coefficients $A_{17,2i+1}$,
	$A_{18,2i+1}$, $A_{19,2i+1}$ and $A_{20,2i+1}$ of the polinomyals
	$v_{17}$, $v_{18}$ $v_{19}$ and $v_{20}$. The coefficients of the
	normalization are $N_{17}=10.704554852432789228$,
	$N_{18}=11.042302467661782828$, $N_{19}=11.371947653621139153$ and
	$N_{20}=11.694019311542212328$.}
\label{tabela12}
\end{table}

\begin{table}
{\tiny\begin{tabular}{|l|l|l|l|}\hline $i$ & Coefficients $A_{21,2i+1}$ & Coefficients $A_{22,2i+1}$ & Coefficients $A_{23,2i+1}$ \\ \hline 
 $0$ & $1$ & $1$ & $1$ \\ \hline 
 $1$ & $-
 58.113569811722124272$ & $- 62.129694188819134863$ & $- 66.218645891923138150
$ \\ \hline 
 $2$ & $ 968.05133093681225903$ & $ 1109.8555369654760655$ & $
 1264.1735619567930545$ \\ \hline 
 $3$ & $- 7313.8405815021465148$ & $-
 9023.3742844288512616$ & $- 11017.892411056715348$ \\ \hline 
 $4$ & $
 30592.495945883842914$ & $ 40777.824146671337701$ & $ 53561.494870460181802
$ \\ \hline 
 $5$ & $- 79175.520459550748409$ & $- 114547.41165151219911$ & $
- 162494.05150986880322$ \\ \hline 
 $6$ & $ 135962.25131813175913$ & $
 214649.82636538022037$ & $ 330362.32813557261476$ \\
\hline \end{tabular}}  
\mycaption{The case $\Lambda=-0.1$, $k=0$: coefficients $A_{21,2i+1}$,
	$A_{22,2i+1}$ and $A_{23,2i+1}$ of the polinomyals $v_{21}$, $v_{22}$
	and $v_{23}$. The coefficients of the normalization are
	$N_{21}=12.009134872819683019$, $N_{22}=12.317752618945874001$ and
	$N_{23}=12.619905741748600855$.}
\label{tabela13}
\end{table}

\begin{table}
{\tiny\begin{tabular}{|l|l|l|l|l|}\hline $i$ & Coefficients $A_{1,2i+1}$ & Coefficients $A_{2,2i+1}$ & Coefficients $A_{3,2i+1}$ & Coefficients $A_{4,2i+1}$ \\ \hline 
 $0$ & $1$ & $1$ & $1$ & $1$ \\ \hline 
 $1$ & $
 21.8377842950992$ & $ 18.4334293031812$ & $ 15.1810746656932$ & $
 12.0920975305193$ \\ \hline 
 $2$ & $ 147.651955803377$ & $
 105.526579144947$ & $ 71.7770440011361$ & $ 45.5993383732682
$ \\ \hline 
 $3$ & $ 483.861595247111$ & $ 292.175608570147$ & $
 163.214861374083$ & $ 81.5591541074805$ \\ \hline 
 $4$ & $
 931.239526525682$ & $ 471.973074791482$ & $ 213.571689647027$ & $
 81.2934444080791$ \\ \hline 
 $5$ & $ 1170.31529375274$ & $
 492.198865923421$ & $ 176.051457215035$ & $ 47.8082621918444
$ \\ \hline 
 $6$ & $ 1026.20921841086$ & $ 351.941764334297$ & $
 95.3901450241527$ & $ 15.8144829846098$ \\
\hline \end{tabular}} 
\mycaption{The case $\Lambda=-0.1$, $k=-1$: coefficients $A_{1,2i+1}$, $A_{2,2i+1}$,
	$A_{3,2i+1}$ and $A_{4,2i+1}$ of the polinomyals $v_{1}$,
	$v_{2}$, $v_{3}$ and $v_{4}$. The coefficients of the
	normalization are   $N_{1}=0.107994257450045 \times 10^{-6}$, 
	$N_{2}=0.292690588075209 \times 10^{-5}$, 
	$N_{3}=0.518252990768902 \times 10^{-4}$ and
	$N_{4}=0.683649822517353\times 10^{-3}$.}
\label{tabela14}
\end{table}
     
\begin{table}
{\tiny\begin{tabular}{|l|l|l|l|l|}\hline $i$ & Coefficients $A_{5,2i+1}$ & Coefficients $A_{6,2i+1}$ & Coefficients $A_{7,2i+1}$ & Coefficients $A_{8,2i+1}$ \\ \hline 
 $0$ & $1$ & $1$ & $1$ & $1$ \\ \hline 
 $1$ & $
 9.18190533268822$ & $ 6.47301073960000$ & $ 4.00115813990478$ & $
 1.80436900132522$ \\ \hline 
 $2$ & $ 26.1743762082396$ & $
 12.6594902235932$ & $ 4.16903822033216$ & $- 0.299804601981924
$ \\ \hline 
 $3$ & $ 34.0957320592592$ & $ 9.92492684765362$ & $
 0.26293559391319$ & $- 1.64081279306699$ \\ \hline 
 $4$ & $
 22.4752715608062$ & $ 2.00505672302545$ & $- 1.74796349658398$ & $-
 0.510193340358966$ \\ \hline 
 $5$ & $ 6.37807566817418$ & $-
 1.67761640342591$ & $- 0.827957567330092$ & $ 0.306140893014826
$ \\ \hline 
 $6$ & $- 0.805445298790648$ & $- 1.11184855823729$ & $
 0.0481471210374753$ & $ 0.175238091233888$ \\
\hline \end{tabular}} 
\mycaption{The case $\Lambda=-0.1$, $k=-1$: coefficients $A_{5,2i+1}$, $A_{6,2i+1}$,
	$A_{7,2i+1}$ and $A_{8,2i+1}$ of the polinomyals $v_{5}$,
	$v_{6}$, $v_{7}$ and $v_{8}$. The coefficients of the
	normalization are   
	$N_{5}=0.699677173459493 \times 10^{-2}$, 
	$N_{6}=0.556647587279093\times 10^{-1}$, 
	$N_{7}=0.330418318317813$ and
	$N_{8}=1.29283805747619$.}
\label{tabela15}
\end{table}

\begin{table}
{\tiny\begin{tabular}{|l|l|l|l|l|}\hline $ l$ & Coefficients $A_{9,2i+1}$ & Coefficients $A_{10,2i+1}$ & Coefficients $A_{11,2i+1}$ & Coefficients $A_{12,2i+1}$ \\ \hline 
 $0$ & $1$ & $1$ & $1$ & $1$ \\ \hline 
 $1$ & $-
 0.232010916759383$ & $- 2.43487417097320$ & $- 4.93455597590118$ & $-
 7.69645852094990$ \\ \hline 
 $2$ & $- 1.85623104021040$ & $-
 0.73836052792331$ & $ 4.05659416695777$ & $ 13.7141424722470
$ \\ \hline 
 $3$ & $- 0.265450448980053$ & $ 2.04821420478302$ & $
 2.31824860373187$ & $- 4.94408507267326$ \\ \hline 
 $4$ & $
 0.786253245370039$ & $ 0.420288826610130$ & $- 2.23653053543589$ & $-
 4.28380968001096$ \\ \hline 
 $5$ & $ 0.232581678867701$ & $-
 0.546742403426884$ & $- 0.530835364516564$ & $ 2.28767539275548
$ \\ \hline 
 $6$ & $- 0.107274956602857$ & $- 0.130901720787139$ & $
 0.424516857735247$ & $ 0.562747928024902$ \\
\hline \end{tabular}} 
\mycaption{The case $\Lambda=-0.1$, $k=-1$: coefficients $A_{9,2i+1}$, $A_{10,2i+1}$,
	$A_{11,2i+1}$ and $A_{12,2i+1}$ of the polinomyals $v_{9}$,
	$v_{10}$, $v_{11}$ and $v_{12}$. The coefficients of the
	normalization are   
	$N_{9}=2.92703125957288$, 
	$N_{10}=4.35635774997061$, 
	$N_{11}=5.39681832616858$ and
	$N_{12}=6.24220211798654$.}
\label{tabela16}
\end{table}

\begin{table}
{\tiny\begin{tabular}{|l|l|l|l|l|}\hline $ i$ & Coefficients $A_{13,2i+1}$ & Coefficients $A_{14,2i+1}$ & Coefficients $A_{15,2i+1}$ & Coefficients $A_{16,2i+1}$ \\ \hline 
 $0$ & $1$ & $1$ & $1$ & $1$ \\ \hline 
 $1$ & $-
 10.6841664013405$ & $- 13.8742150817412$ & $- 17.2503270722203$ & $-
 20.7999881540638$ \\ \hline 
 $2$ & $ 29.3147115073451$ & $
 51.8840233684938$ & $ 82.4201447999443$ & $ 121.901220187412
$ \\ \hline 
 $3$ & $- 27.8107960004119$ & $- 76.9972681817258$ & $-
 165.993521308318$ & $- 311.189756974583$ \\ \hline 
 $4$ & $
 4.6218685711793$ & $ 46.6058507383930$ & $ 160.587831324991$ & $ 407.968109284316
$ \\ \hline 
 $5$ & $ 6.17023638414212$ & $- 3.5351498009348$ & $-
 69.8936934649346$ & $- 287.865946240682$ \\ \hline 
 $6$ & $-
 2.25304381758315$ & $- 7.59830396617375$ & $ 2.86651578722247$ & $
 99.0775488526631$ \\
\hline \end{tabular}} 
\mycaption{The case $\Lambda=-0.1$, $k=-1$: coefficients $A_{13,2i+1}$, $A_{14,2i+1}$,
	$A_{15,2i+1}$ and $A_{16,2i+1}$ of the polinomyals $v_{13}$,
	$v_{14}$, $v_{15}$ and $v_{16}$. The coefficients of the
	normalization are   
	$N_{13}=6.98413261507519$, 
	$N_{14}=7.65951898965145$, 
	$N_{15}=8.28713472086337$ and
	$N_{16}=8.87811017222452$.}
\label{tabela17}
\end{table}

\begin{table}
{\tiny\begin{tabular}{|l|l|l|}\hline $i$ & Coefficients $A_{17,2i+1}$ & Coefficients $A_{18,2i+1}$ \\ \hline 
 $0$ & $1$ & $1$ \\ \hline 
 $1$ & $- 24.5135934862227$ & $-
 28.3826830252342$ \\ \hline 
 $2$ & $ 171.297635578830$ & $
 231.563657333425$ \\ \hline 
 $3$ & $- 532.032302271353$ & $-
 851.104089577650$ \\ \hline 
 $4$ & $ 879.049849725663$ & $
 1700.00285906748$ \\ \hline 
 $5$ & $- 837.566899410330$ & $-
 2028.92845736494$ \\ \hline 
 $6$ & $ 466.195823019623$ & $
 1512.52404634436$ \\ \hline 
 $7$ & $- 137.342049715716$ & $-
 704.222275510342$ \\ \hline 
 $8$ & $ 8.59568426834457$ & $
 188.892562917903$ \\ \hline 
 $9$ & $ 7.36166712174031$ & $-
 17.5315261277321$ \\ \hline 
 $10$ & $- 2.12487622903933$ & $-
 5.41896929313905$ \\ \hline 
 $11$ & $ 0.0616353109016188$ & $
 1.93593317266994$ \\ \hline 
 $12$ & $ 0.0612186597075610$ & $-
 0.163977083115011$ \\ \hline 
 $13$ & $- 0.00828671964266124$ & $-
 0.0263879773884429$ \\ \hline 
 $14$ & $- 0.00051214720454678$ & $
 0.00637464986776314$ \\ \hline 
 $15$ & $ 0.000167433511984800$ & $-
 0.000186792886772932$ \\ \hline 
 $16$ & $- 0.00000303760478624729$ & $-
 0.0000646259429141273$ \\ \hline 
 $17$ & $- 0.00000147924868677497$ & $
 0.00000596587830124041$ \\ \hline 
 $18$ & $ 0.000000075304283408867
$ & $ 0.000000160392422977485$ \\ \hline 
 $19$ & $
 0.00000000606586647233395$ & $- 0.0000000400489171784274
$ \\ \hline 
 $20$ & $- 0.000000000434054633266702$ & $
 0.00000000080165107135159$ \\ \hline 
 $21$ & $-{ 9.3095562785296
\times 10^{-12}}$ & ${ 8.54942653394270\times 10^{-11}}
$ \\ \hline 
 $22$ & ${ 8.32964788661547\times 10^{-13}}$ & $-{
 3.37515341995877\times 10^{-12}}$ \\ \hline \end{tabular}} 
\mycaption{The case $\Lambda=-0.1$, $k=-1$: coefficients $A_{17,2i+1}$ and $A_{18,2i+1}$
	of the polinomyals $v_{17}$ and $v_{18}$. The coefficients of
	the normalization are   
	$N_{17}=9.43144441465169$ and 
	$N_{18}=9.75064661119335$.}
\label{tabela18}
\end{table}
\newpage

\end{document}